\documentclass[aps, superscriptaddress ,twocolumn,prl]{revtex4}
\usepackage{amsmath}
\usepackage{amssymb}
\usepackage{epsfig}
\usepackage{graphics}
\begin{document}
\title{Shear viscosity and spin diffusion coefficient of a two-dimensional Fermi gas}
\author{G.\ M.\ Bruun}
\affiliation{Department of Physics and Astronomy, University of Aarhus, Ny Munkegade, 8000 Aarhus C, Denmark}
\begin{abstract}
Using kinetic theory, we calculate the shear viscosity and the spin diffusion coefficient as well as  the associated relaxation times 
for a two-component Fermi gas in two dimensions, as a function of temperature, coupling strength, polarization, and 
mass ratio of the two components.  It is demonstrated that the minimum value 
of the viscosity decreases  with the mass ratio, since Fermi blocking becomes less efficient.
 We furthermore analyze recent experimental 
results for the quadrupole mode of a 2D gas in terms of viscous damping obtaining a qualitative agreement using no fitting parameters.

\end{abstract}
\maketitle

\section{Introduction}
The properties of 2-dimensional (2D) Fermi systems are fundamental for  our understanding 
of a wide range of  phenomena  including organic and high-$T_c$ superconductors, 2D nano-structures, and $^3$He films. 
A new generation of  experiments are now probing the many-body properties of atomic Fermi gases in 2D traps~\cite{Martiyanov,Feld,Sommer}.
This provides a unique possibility to systematically explore the physics of 2D systems using the high experimental control characterizing 
atomic gases. Recently, there has been a lot of interest in the transport properties of atomic gases. 
One reason is that transport coefficients provide excellent probes for strong correlations, since they can change by orders of magnitude due to interactions.
It has been shown experimentally  that 3D
atomic gases may form a perfect fluid with a shear viscosity $\eta$ having the least possible value consistent with quantum mechanics~\cite{Cao}.
This has inspired a lot work investigating the connections between the physics of atomic gases, and other strong coupling systems including quark-gluon plasmas, and  
liquid Helium~\cite{SchaferTeaney}. 
Also, recent experiments  demonstrate that the spin diffusion coefficient approaches a scale set by quantum mechanics  
for a 3D resonantly interacting atomic gas~\cite{SommerSpinDiff}. 
The first experiments probing the collective mode spectrum of a strongly interacting 2D Fermi gas were recently reported~\cite{Vogt}. The frequency  of the breathing mode 
was shown to be provide evidence of a classical dynamical scaling symmetry~\cite{Hofmann}, whereas the damping of the quadrupole mode was used as a measure 
for the shear viscosity of a 2D Fermi gas. 

We calculate the shear viscosity and the spin diffusion coefficient as well as the associated relaxation times 
for a  two-component Fermi gas in 2D using kinetic theory. The dependence of the viscosity on the mass ratio of the two components is analyzed, 
and we show that the minimum value is reduced for systems with a mass imbalance.  
We furthermore analyze the recent experimental results for the quadrupole mode in 
terms of viscous damping~\cite{Vogt} obtaining a qualitative agreement. However, our analysis shows that further work is needed 
to understand the experiments quantitatively.

\section{Formalism}
Consider a 2D gas of  two fermionic species $\sigma=1,2$ with mass $m_\sigma$ 
and density $n_\sigma= k_{F\sigma}^2/4\pi$ so that the total density is $n=n_1+n_2$. The 
range of the interaction is taken to be much shorter than the interparticle spacing, and there is therefore no interaction 
between identical fermions. We shall focus on two steady state 
non-equilibrium situations: one with a spatially varying local mean velocity  ${\mathbf u}({\mathbf r})$,
and one with a spatially varying  magnetization  $M({\mathbf r})=n_1({\mathbf r})-n_2({\mathbf r})$, 
which we for concreteness take to have the forms ${\mathbf u}({\mathbf r})=[u_x(y),0]$ and  $M({\mathbf r})=M(x)$.
As a result of the velocity field $u_x( y)$, there is a net current $\Pi_{xy}$ in the $y$-direction of momentum along the $x$-direction, 
and likewise $M(x)$ induces a net magnetization current $j_M$ in the $x$-direction. Within linear response, we can write 
\begin{align}
\Pi_{xy}=-\eta\partial_y u_x&&\text{and}&& j_M=-D\partial_x M
\label{definition}
\end{align}
which  defines the shear viscosity $\eta$ and the spin diffusion coefficient $D$. 

We briefly outline a variational method  to calculate the shear viscosity  and the spin diffusion coefficient  within kinetic theory. Further details are 
given in Refs.~\cite{SmithHojgaard,BaymPethick,MBS,GMBNJP}. 
In kinetic theory, both coefficients are obtained from a steady state solution to the Boltzmann equation. 
In the hydrodynamic limit, the distribution functions  $f_\sigma({\mathbf r},{\mathbf p})$ are close to the local equilibrium form
 $f^{\rm le}_\sigma=1/[\exp(\beta\xi^{\rm le}_{\sigma})+1]$ with 
$\xi^{\rm le}_{\sigma}=p^2/2m_\sigma-{\mathbf u}({\mathbf{r}})\cdot{\mathbf p} -\mu_\sigma$ for the case of a local 
velocity field ${\mathbf u}$ and 
$\xi^{\rm le}_{\sigma}=p^2/2m_\sigma -\mu_\sigma({\mathbf{r}})$ appropriate 
for a local magnetization $M$. Here  $\mu_1({\mathbf{r}})$ and  $\mu_2({\mathbf{r}})$ are the 
spatially varying chemical potentials corresponding to the magnetization and $\beta=1/T$ (we use units where $k_B=\hbar=1$).
When these local equilibrium functions are plugged into the left side of the linearized Boltzmann equation, it 
can be written as 
\begin{align}
\frac{\partial f^0_{\sigma}}{\partial \epsilon}\Phi_{\sigma}^\eta\frac{\partial u_x}{\partial y}=I_\sigma &&\text{and}&& 
\frac{\partial f^0_{\sigma}}{\partial \epsilon}\Phi_{\sigma}^D\frac{\partial\mu_\sigma}{\partial x}=I_\sigma
\label{BE}
\end{align}
where $I_\sigma$ is the collision operator for component $\sigma$, and $f^0_\sigma=1/[\exp\beta(\epsilon_\sigma-\mu_\sigma)+1]$
 is the equilibrium function with $\epsilon_\sigma=p^2/2m_\sigma$. Here $\Phi_\sigma^\eta=p_xp_y/m_\sigma$ 
and $\Phi_\sigma^D=p_x/m_\sigma$ for shear viscosity and spin diffusion respectively.
The momentum and spin currents are given by 
\begin{gather}
\Pi_{xy}=\int d^2\check{k} (\Phi_1^\eta f_1+ \Phi_2^\eta f_2)\nonumber\\
 j_{M}=\int d^2\check{k}(\Phi_1^Df_1-\Phi_2^Df_2)
 \label{currents}
 \end{gather}
 with $d^2\check{k}=d^2k/(2\pi)^2$. To proceed, we need an approximate solution to the Boltzmann equation (\ref{BE}).
 In 3D, the ansatz $\delta f_\sigma\propto \Phi_\sigma f^0_\sigma(1-f^0_\sigma)$ for 
 the deviation of $f_\sigma({\mathbf r},{\mathbf p})$ away from equilibrium 
is known to yield results within $2\%$ of the exact result for the viscosity~\cite{SmithHojgaard,BruunSmith}. We therefore use this ansatz for the 2D case
which   yields 
\begin{align}
\eta=2\beta\frac{\langle {\Phi^\eta}^2\rangle^2}{\langle \Phi^\eta H[\Phi^\eta]\rangle}&&\text{and}&&
D=\frac{\beta}{\chi}\frac{\langle {\Phi^D}^2\rangle^2}{\langle \Phi^D H[\Phi^D]\rangle}.
\label{etaogD}
\end{align} 
as  variational expressions for the viscosity and spin diffusion coefficient. 
We have defined the average 
$\langle \Phi^2 \rangle\equiv2^{-1}\sum_\sigma\int d^2\check{k}f^0_\sigma(1-f^0_\sigma)\Phi_\sigma^2$
 and $\chi=\partial (n_1-n_2)/\partial(\mu_1-\mu_2)$ is
 the magnetic susceptibility. The linearized collision integral can after symmetrization be written as 
\begin{gather}
\langle \Phi H[\Phi]\rangle=
\frac{1}{4}\int d^2\check{k}_1d^2\check{k}_2\frac{p_r}{m_r}\int_0^{2\pi}d\theta'\frac{d\sigma}{d\theta'}\nonumber\\
\times(\Delta\Phi)^2f^0_1 f^0_2(1-f^0_1)(1-f^0_2)
\label{CollOp}
\end{gather} 
where  ${\mathbf p}_r=(m_2{\mathbf p}_1-m_1{\mathbf p}_2)/M$ is the relative momentum of the incoming scattering 
particles,  $\theta'$ is the angle between the outgoing and incoming relative momenta, and 
$d\sigma/d\theta$ is the differential cross section. The total mass is $M=m_1+m_2$,  
and $m_r^{-1}=m_1^{-1}+m_2^{-1}$ is the reduced mass. We have defined the function 
 $\Delta\Phi\equiv\Phi_{1}^\eta+\Phi_{2}^\eta-\Phi_{3}^\eta-\Phi_{4}^\eta$
for shear viscosity and $\Delta\Phi\equiv\Phi_{1}^D-\Phi_{2}^D-\Phi_{3}^D+\Phi_{4}^D$
for spin diffusion. It determines the contribution of a given collision to the momentum and spin transport   respectively. 
The reason for the factor $2$ difference  in the expressions for  $\eta$ and $D$ 
in (\ref{etaogD}) is that the two components contribute with the same sign to 
 the momentum current  and with opposite signs to  the
magnetic current, see (\ref{currents}). This also causes the sign differences in the expressions for  $\Delta\Phi$ in the two cases. 
 
 \subsection{Relaxation times}
The  viscous and spin relaxation times $\tau_\eta$ and $\tau_D$ which give the typical 
time between collisions for the two types of motion are useful for estimating whether a system is in the hydrodynamic regime. 
  Suitable definitions can be obtained by writing the 
 collision integral  as $I_\sigma\simeq\delta f_\sigma/\tau$.
   which gives 
 $\eta=2\tau_\eta\beta\langle (\Phi^\eta)^2\rangle$ and $D=\tau_D\beta\langle (\Phi^D)^2\rangle/\chi$.
 Performing the integrals yields
 \begin{align}
\frac{\eta}{n}=\tau_\eta\frac{\int_0^\infty d\epsilon \epsilon f^0}{\int_0^\infty d\epsilon  f^0}&&\text{and}&&\frac{D}{n}=
\frac{\tau_D}{2m\chi}
\label{tau}
\end{align}
where we have taken  $m_1=m_2$  and $n_1=n_2=n/2$ for simplicity. 
This reduces to $\eta=n\tau_\eta\epsilon_F/2$ and $D=\epsilon_F\tau_D/m$ in the  degenerate  limit, whereas 
 $\eta=n\tau_\eta T$ and $D=\tau_D T/m$ in the  classical limit. We have used 
 $\chi=m/2\pi$ for $T\ll T_F$ and $\chi=n/2T$ for $T\gg T_F$.

 \subsection{Scattering cross section}
When the range of the interaction is much shorter than the typical interparticle spacing, the
 scattering between the $\sigma=1$ and $\sigma=2$  fermions  is predominantly $s$-wave.  The 2D cross section for relative momentum $p_r$ is $\sigma=m_r^2|{\mathcal T}(p_r^2/2m_r)|^2/p_r$
with  the  ${\mathcal T}$-matrix given by~\cite{Adhikari,Petrov,Mohit}
\begin{equation}
{\mathcal T}(\epsilon)=\frac{2\pi}{m_r}\frac{1}{\ln(|E_b|/\epsilon)+i\pi}
\label{Tmatrix}
\end{equation}
which has a pole at a 2-body bound state with energy $E_b=-1/2m_ra_2^2$.

\section{Classical limit}
Consider the classical limit $T\gg T_{F\sigma}=k_{F\sigma}^2/2m_\sigma$.
 In this limit $f^0_\sigma\ll 1$, and the integrals in (\ref{etaogD}) are straightforward to perform. 
We obtain for the viscosity 
\begin{equation}
\eta_{\rm cl}=\frac{m_r}{2\pi^2}\frac{(n_1+n_2)^2}{n_1 n_2}\frac{T}{I_\eta(T/E_b)}
\label{etaclass}
\end{equation}
with 
\begin{gather}
I_\eta(T/E_b)=\int_0^{\infty}dte^{-t}\frac{t^2}{\ln(|E_b|/Tt)^2+\pi^2}\nonumber\\
\simeq\frac{2}{[\ln(|E_b|/T)-0.92]^2+\pi^2}.
\label{Ieta}
\end{gather}
The shear viscosity  depends only on the reduced mass in the classical limit. This is because the scattering 
only depends on the relative coordinates in this limit, 
 since there is no Fermi-blocking. 
For fixed total density $n_1+n_2$, the viscosity is minimum for 
$n_1=n_2$ as expected, since the scattering becomes less frequent with increasing population imbalance. 
The viscosity in the classical limit for  $m_1=m_2$ and $n_1=n_2$ was reported 
while this manuscript was being written~\cite{Schafer} and the result agrees with  (\ref{etaclass})-(\ref{Ieta}) for that  case. 

Likewise, (\ref{etaogD}) yields for the spin diffusion coefficient in the classical limit 
\begin{equation}
D_{\rm cl}=\frac{T}{4\pi^2 n}\frac{1}{I_D(T/E_b)}
\label{Dclass}
\end{equation}
with 
\begin{gather}
I_D(T/E_b)=\int_0^{\infty}dte^{-t}\frac{t}{\ln(|E_b|/Tt)^2+\pi^2}\nonumber\\
\simeq\frac{1}{[\ln(|E_b|/T)-0.42]^2+\pi^2}.
\label{ID}
\end{gather}
Here, we have for simplicity taken $m_1=m_2$ and $n_1=n_2=n/2$ and used $\chi=n/2k_BT$
in the classical limit. 

Equations (\ref{etaclass})-(\ref{Ieta})  and (\ref{Dclass})-(\ref{ID}) should be compared with the analogous expressions obtained 
for the 3D case: $\eta= 15(mk_BT)^{3/2}/32\sqrt\pi$~\cite{MBS,BruunSmith2} and $D=3\sqrt m (k_BT)^{3/2}/16\sqrt\pi$~\cite{GMBNJP}
for a classical gas in the unitarity regime. The reason for the more complicated $T$-dependence in 2D is  the intrinsic energy 
dependence of the ${\mathcal T}$ matrix  (\ref{Tmatrix}), which means
one never recovers the simple power law predictions for an energy independent cross section:
$\eta\sim nvml_{\rm mf}\propto\sqrt T$ and $D\sim vl_{\rm mf}\propto\sqrt T$ where $l_{\rm mf}\sim1/n\sigma$ is 
the mean free path.

\section{Numerical results}
In this section we present numerical results for the viscosity and spin diffusion coefficient obtained from (\ref{etaogD}).
\subsection{Viscosity}
\begin{figure}
\includegraphics[width=1\columnwidth]{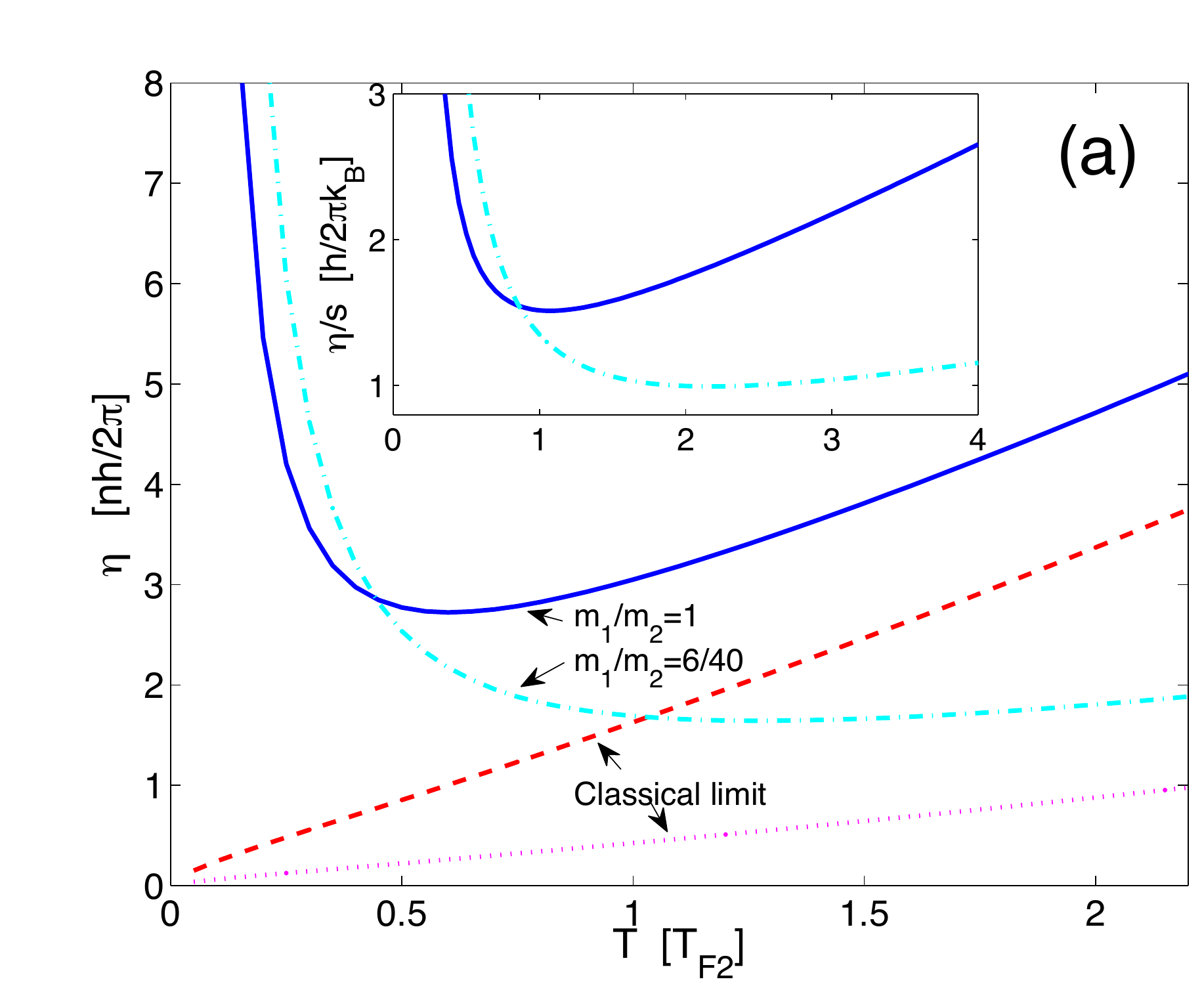}
\includegraphics[width=1\columnwidth]{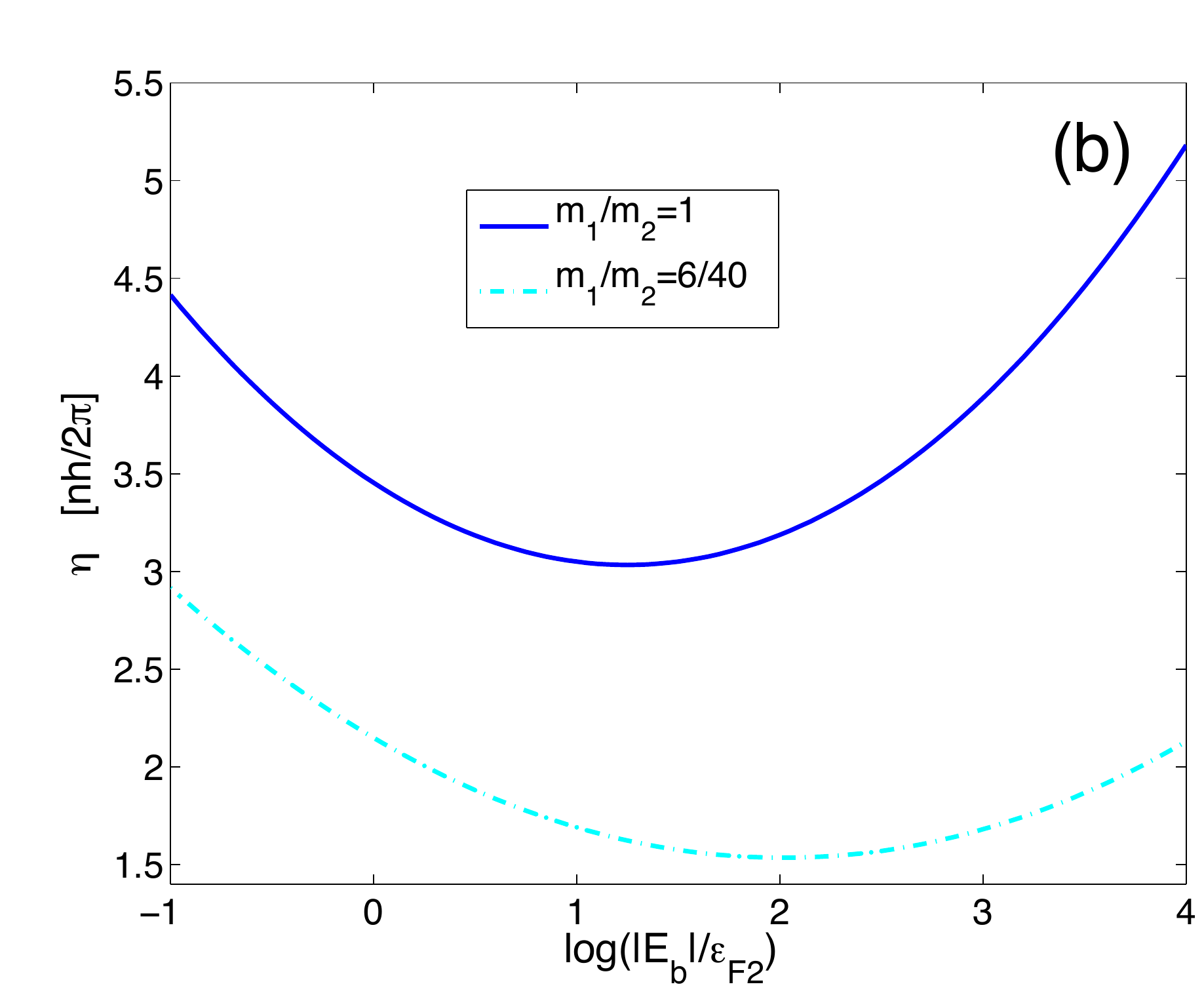}
\caption{(color on-line)(a) The viscosity as a function of temperature for $\ln(|E_b|/T_{F2})=1$ and $m_1/m_2=1$ (blue line), 
$m_1/m_2=6/40$ (cyan dash-dot line). The classical limits are plotted as red dashed and purple dotted lines.  
The inset shows $\eta/s$ as a function of $T$.
(b)  The viscosity as a function of $\ln(|E_b|/T_{F2})$ for $T=T_{F2}$ and $m_1/m_2=1$ (blue  line), and
$m_1/m_2=6/40$ (cyan dash-dot line).  
}
\label{etaFig}
\end{figure}

In Fig.~\ref{etaFig} (a), we plot the viscosity
as a function of $T$ for the mass ratios $m_1/m_2=1$ and $m_1/m_2=6/40$. The  
latter corresponds to a  mixture of $^{40}$K and $^6$Li atoms. We have taken the density of the two components to be equal and 
 $\ln(|E_b|/T_{F2})=1$. 
For high temperatures, the viscosity approaches the classical result (\ref{etaclass}), whereas it increases strongly for 
low $T$ due to Fermi blocking~\cite{Novikov}. This results in a minimum of the viscosity at $T\simeq0.6T_{F2}$ 
for $m_1/m_2=1$, whereas the minimum is located at  $T\simeq1.3T_{F2}$  for $m_1/m_2=6/40$ due to the larger Fermi temperature 
for the light component $\sigma=1$. 

An important result is that the minimum viscosity of the mass imbalanced mixture is significantly smaller 
than for the mass balanced mixture; for the mass ratio $m_1/m_2=6/40$ it is a factor $0.6$ smaller. 
This can be understood as follows:  changing both masses keeping $m_1/m_2=1$ clearly does 
not reduce the minimum value of $\eta/n$, since this simply amounts to rescaling $T_F$; however, reducing $m_1$ while keeping $m_2$ fixed
makes the Fermi blocking less efficient on the scale of $T_{F1}$ and the minimum value of $\eta/n$ is reduced essentially 
since the classical result (\ref{etaclass}) holds for lower $T/T_{F1}$. The minimum value of the viscosity is subject to intense interest 
due to a conjecture inspired by results for a certain class of strong coupling theories~\cite{Policastro}, which states that the ratio 
of the viscosity over the entropy of any system obeys the  universal bound $\eta/s>1/4\pi$~\cite{Kovtun}. In 
 the inset of Fig.~\ref{etaFig} (a),  we therefore plot $\eta/s$
for the same parameters as in the main plot. The entropy density $s=s_1+s_2$ is obtained from the ideal gas expression 
$s_\sigma=-\int d^2\check k[f^0_\sigma\ln f_\sigma^0-(1-f^0_\sigma)\ln(1-f_\sigma^0)]$. Again, we see that the minimum value of $\eta/s$ is significantly 
smaller for the mass ratio $m_1/m_2=6/40$. Intriguingly, it seems to follow from kinetic theory 
 that a two-component systems with a sufficiently  large mass ratio 
can break the conjectured   bound $\eta/s>1/4\pi$.  A similar effect is in fact present for 3D systems.

In Fig.~\ref{etaFig} (b), we plot $\eta/n$ as a function of $\ln(|E_b|/T_{F2})$ for $T=T_{F2}$. 
The viscosity is minimum in the strong coupling regime $\ln(|E_b|/T_{F2})\sim{\mathcal{O}}(1)$ as expected. In the classical 
limit, it  follows  from (\ref{etaclass}) that the minimum is located at $\ln(|E_b|/T)\sim 0.92$. With decreasing $m_1/m_2$, 
the minimum moves to larger values of $\ln(|E_b|/T_{F2})$ because $T_{F1}$ increases.

\subsection{Spin diffusion coefficient}
\begin{figure}
\includegraphics[width=1\columnwidth]{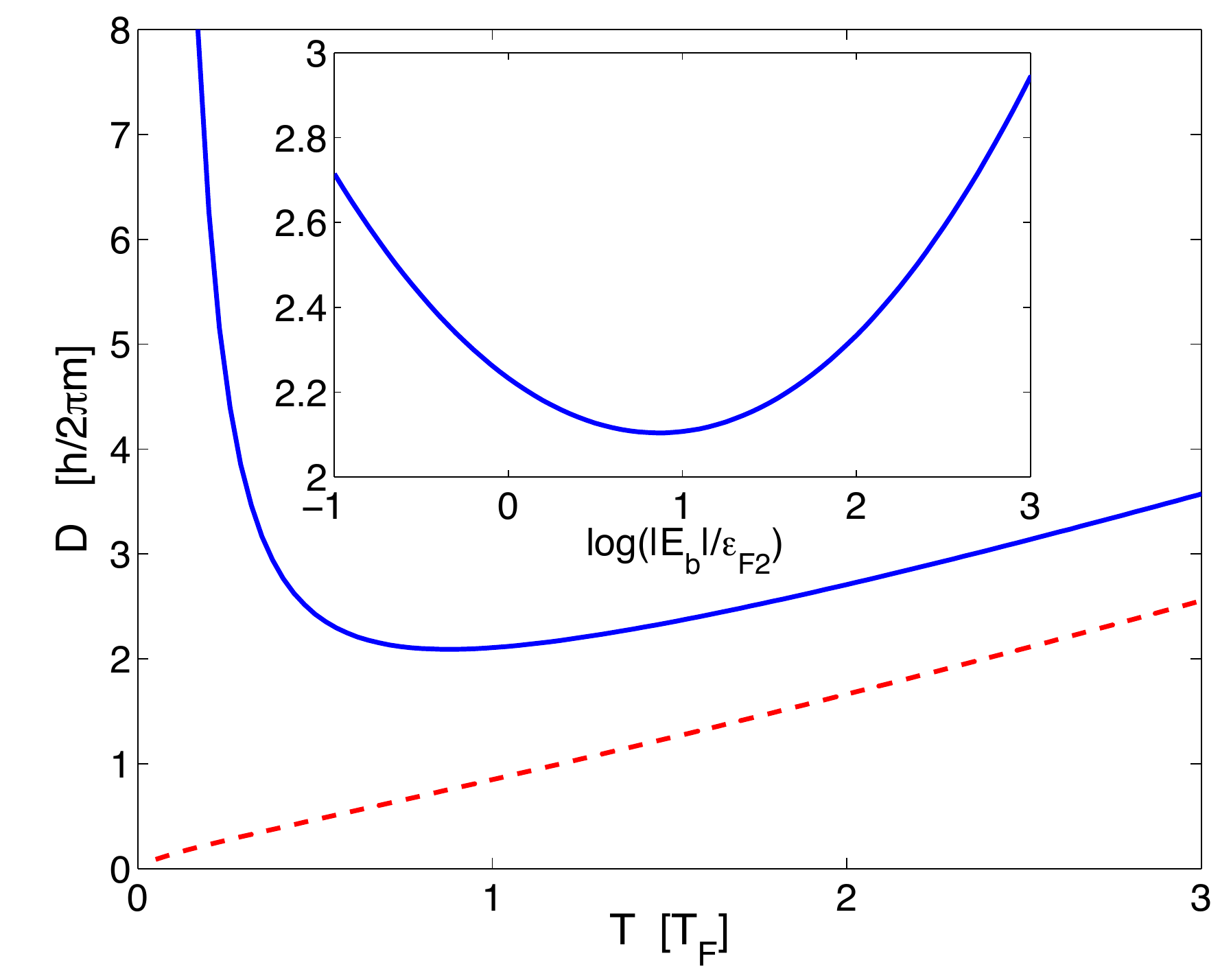}
\caption{(color on-line) The spin diffusion coefficient  as a function of temperature for $\ln(|E_b|/T_{F2})=1$ and $m_1/m_2=1$. 
The classical limit is plotted as a red dashed  line. The inset shows $D$ as a function  of the interaction strength for $T=T_{F2}.$}
\label{DFig}
\end{figure}

The spin diffusion coefficient is plotted in  Fig.~\ref{DFig} as a function of temperature for  $m_1/m_2=1$
and  $\ln(|E_b|/T_{F2})=1$. For high $T$, it approaches the classical value (\ref{Dclass}) whereas Fermi blocking makes it increase strongly for 
low $T$, leading to a minimum value at $T=0.85T_F$.  
The inset shows $D$ as a function of $\ln(|E_b|/T_{F2})$ for $T=T_{F2}$. Again, the minimum value is for $\ln(|E_b|/T_{F2})\sim{\mathcal{O}}(1)$.
For high $T$, (\ref{Dclass}) predicts $\ln(|E_b|/T)\sim 0.42$ to be the  minimum value. 

\subsection{Validity of kinetic theory}
Let us briefly discuss the range of  validity of  kinetic theory. For weak coupling $|\ln(|E_b|/\epsilon_F|)|\gg1 $,  the kinetic approach is accurate except for 
extremely low temperature, as has been shown for the viscosity in 3D~\cite{BruunSmith2}. For strong coupling, one must expect the Boltzmann  equation (\ref{BE}) to break down 
for low temperature where there are no  well-defined quasi-particles. 
In 3D, calculations of the viscosity based on the Kubo-formalism  show that 
the kinetic approach is  accurate down to temperatures significantly below $T_F$, even for strong coupling~\cite{Enss,BruunSmith,BruunSmith2,Riedl}. We 
expect a similar result to hold in 2D. 
In particular, kinetic theory is likely to be reliable at the temperatures where we predict $\eta$ and $D$ to be minimum. 
The occupation of the closed channel molecule can furthermore have significant effects on  thermodynamic properties in 2D~\cite{Duan}.
Similarly, corrections to the single channel approximation for the ${\mathcal T}$-matrix (\ref{Tmatrix}) could influence the 
transport properties considered here,  although one would expect small effects  for a broad resonance.

\section{Experiments}
The frequency and damping of the quadrupole mode of a 2D Fermi gas of $^{40}$K atoms with equal 
populations in two hyperfine states were recently measured~\cite{Vogt}. The results were interpreted in terms of viscous damping 
appropriate for the hydrodynamic regime. The amplitude damping of a collective mode  can be calculated from~\cite{LandauLifshitz} 
 \begin{equation}
 \Gamma=\frac{|\langle\dot{E}_{\rm mech}\rangle_t|}{2\langle E_{\rm mech}\rangle_t} 
 \label{dynhydro}
\end{equation}
where $\langle E_{\rm mech}\rangle_t$ is the time averaged mechanical energy of the mode. Taking the velocity field of the quadrupole mode to have   
the form ${\mathbf u}({\bf r})=(x,-y)\cos \omega t$,  we get 
\begin{equation}
 \langle E_{\rm mech}\rangle_t=\frac{m}{2} \int d^2{r}\,n({\bf
r})u^2({\bf r}),\label{Emec}
\end{equation}
where we have used that the potential energy of the mode is equal to the kinetic energy, and we have neglected any interaction energy. 
The viscous damping is for this velocity field, following \cite{LandauLifshitz,BruunSmithScissor},  given by 
\begin{equation}
\langle\dot{E}_{\rm mech}\rangle=-2b^2\int d^2{ r}\,\frac{\eta}{1+\omega_Q^2\tau_\eta({\bf r})^2}. \label{rateofloss}
\end{equation}
Here we have used  the real part of the complex dynamical viscosity
$\eta(\omega)=\eta/[1-i\omega\tau_\eta({\bf r})]$~\cite{Nikuni} evaluated at the quadrupole frequency $\omega_Q$
 to obtain a cut-off in the outer classical regions of the cloud, where the viscosity 
is given by  (\ref{etaclass}) with $m_1=m_2$ and $n_1=n_2=n/2$, 
and therefore is independent of density. In the classical regime, (\ref{dynhydro}) becomes using (\ref{etaclass}) and (\ref{tau})
 \begin{equation}
 \Gamma_{\rm cl}=\frac{2\omega}{\pi\sqrt{N}I_\eta}\int_0^\infty du\frac{1}{1+\omega_Q^2\tau_{\eta}^2}
 \label{dynhydroclass}
\end{equation}
where $N=N_1+N_2$ is the total number of particles trapped and  $\omega$ is the 2D trapping frequency.

\begin{figure}
\includegraphics[clip=true,width=0.98\columnwidth,height=1\columnwidth]{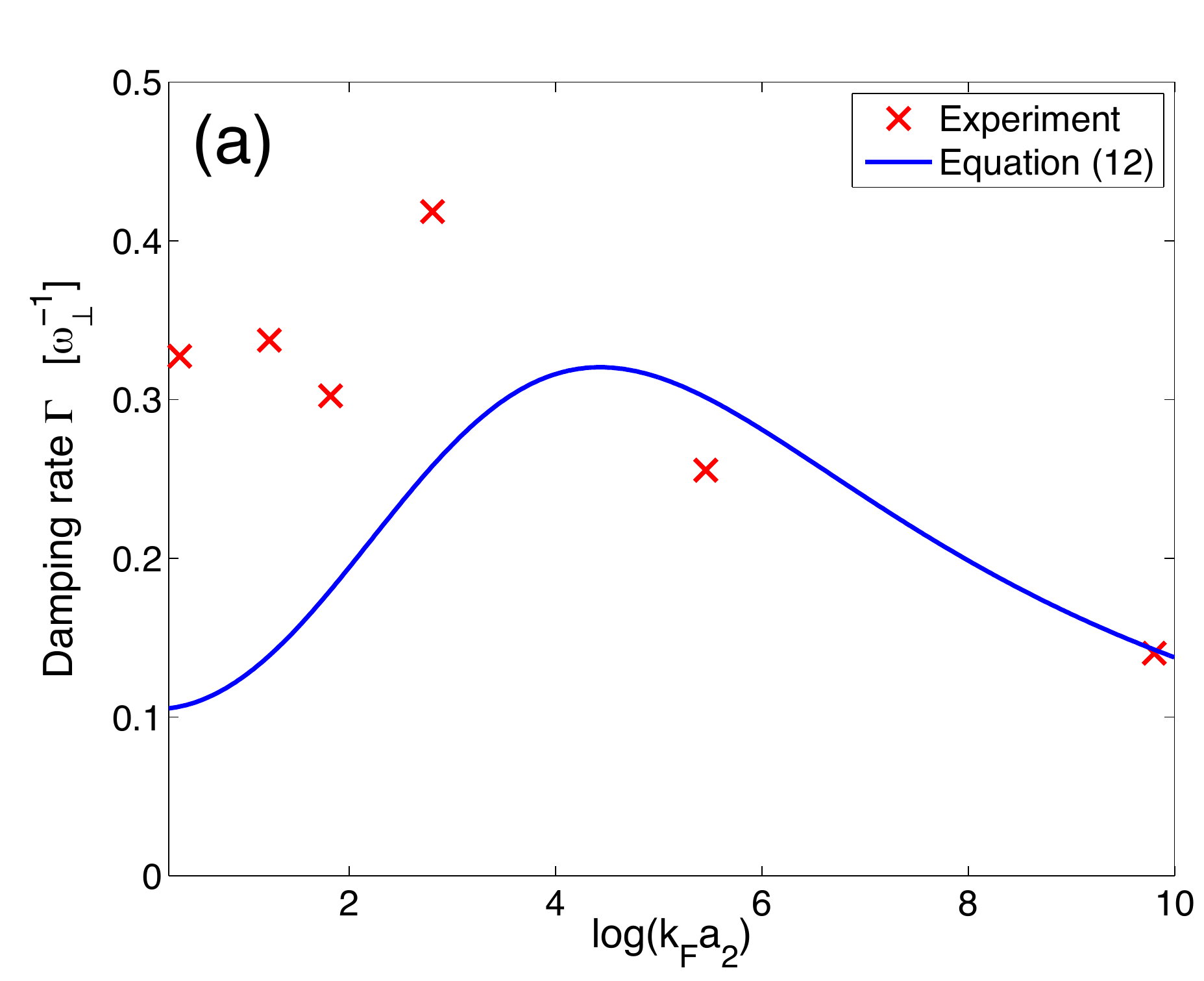}
\includegraphics[clip=true,width=0.98\columnwidth,height=1\columnwidth]{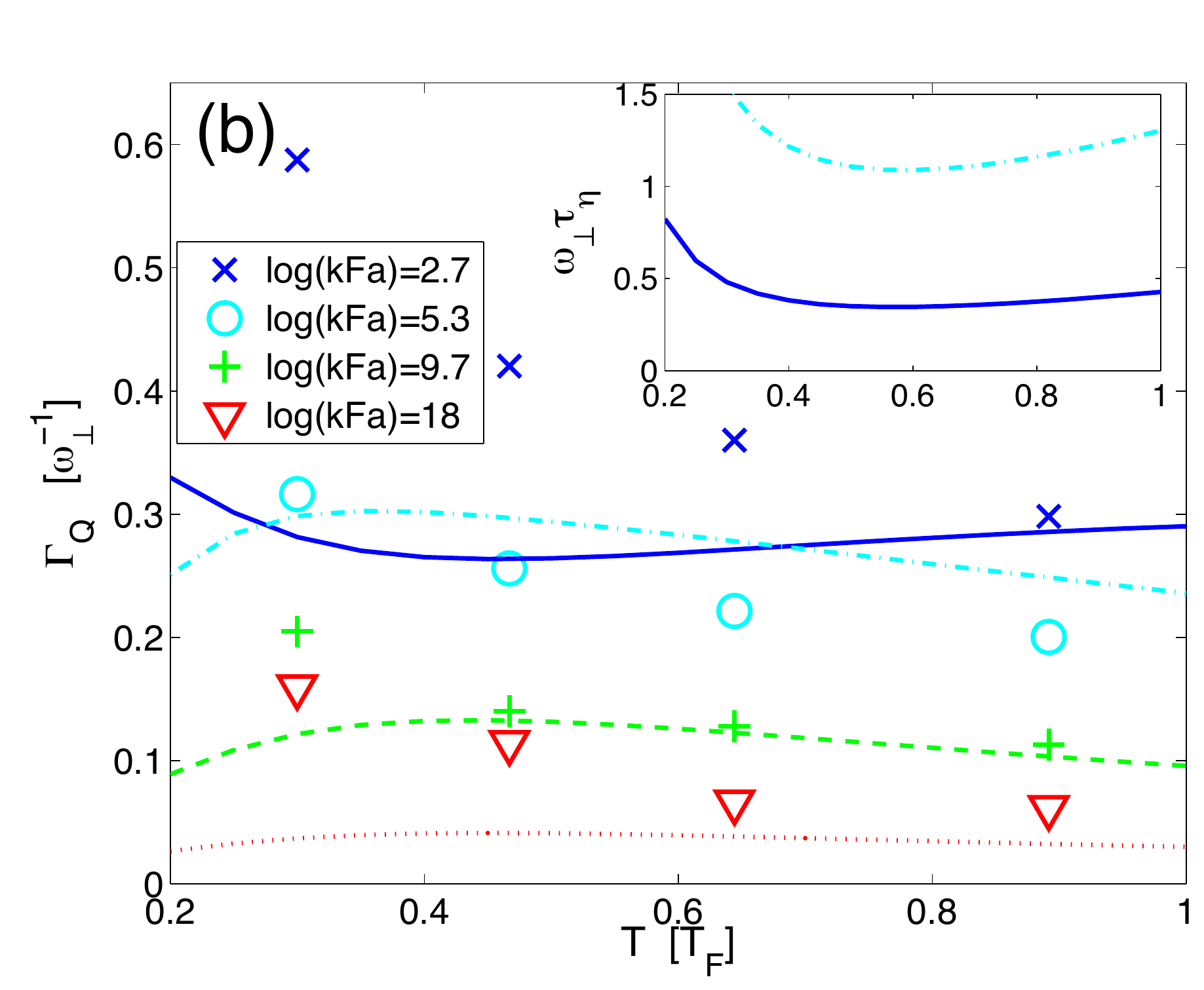}
\caption{(color on-line) (a) The damping of the quadrupole mode as a function of interaction strength. The $\times$'s are 
the experimental results of Ref.~\cite{Vogt}.
(b)  The damping of the quadrupole mode as a function of $T$ for various coupling strengths with the $\times$'s  
the experimental results of Ref.~\cite{Vogt}. The inset shows the viscous collision rate $\omega_\perp\tau_\eta$.}
\label{ExpFig}
\end{figure}

In Fig.~\ref{ExpFig} (a), we plot the damping of the quadrupole mode taking $T/T_F=0.47$ and $N=(E_F/\hbar\omega_\perp)^2=4300$
 particles trapped which are the experimental parameters appropriate for Fig.\ 1 in  Ref.~\cite{Vogt}.
We have calculated the damping  as a function of $\ln(k_Fa_2)$ and the $\times$'s are the experimental results 
reported in Ref.~\cite{Vogt}. We see that the theory  agrees  qualitatively with the data.  In Fig.~\ref{ExpFig} (b), we plot the damping as 
a function of $T$ for various coupling strengths  taking $N=3500$ particles trapped to model the experimental 
situation of Fig.~3 in Ref.~\cite{Vogt}.  Again, the theory accounts qualitatively for the 
experimental results which are plotted as $\times$'s. Note that we have no fitting parameters. For both sets of data, the agreement between theory and experiment 
is best for large $\ln(k_Fa_2)$ and large $T/T_F$ whereas there are significant quantitative discrepancies 
in the strong coupling regime of small $\ln(k_Fa_2)$. Similar results were reported in Ref.~\cite{Schafer}  
for the spatial average of the viscosity using the classical limit approximation.  

It is perhaps surprising 
that the agreement is best in the weak coupling regime where the system is collisionless rather than hydrodynamic.
 This is illustrated in the inset in Fig.~3 (b) which shows $\omega_\perp\tau_\eta$:
the hydrodynamic condition $\omega_\perp\tau_\eta<1$ is fulfilled  only for 
$\ln(k_Fa_2)=2.7$, where as $\omega_\perp\tau_\eta>1$ for larger $\ln(k_Fa_2)$ indicating collisionless dynamics. 
However, despite being based on hydrodynamics, the viscous damping approach turns out to  work rather well in the 
collisionless regime. In 3D it has in fact been shown to yield exact results in the collisionless limit provided one uses the 
complex dynamical viscosity evaluated at the collisionless collective mode frequency~\cite{BruunSmithScissor}.

 The reason for the discrepancy 
between theory and experiment for small $\ln(k_Fa_2)$ where the system is  hydrodynamic, can be strong coupling effects making the kinetic approach 
quantitatively inaccurate as discussed above. 
 Better agreement could also be obtained by solving the Boltzmann equation approximately by taking moments with basis functions for $\delta f_\sigma$~\cite{BruunSmithScissor,Riedl,Chiacchiera}. Such an approach  has indeed been  successful in describing the 
frequency and damping of collective modes in 3D.

To summarize, we have, using kinetic theory, calculated the shear viscosity and the spin diffusion coefficient for a two-component Fermi gas in 2D.
Both transport coefficients have a minimum value somewhat below the Fermi temperature. We showed that the
minimum value of the viscosity can be reduced significantly with increasing mass ratio of the two components.  Using a viscous damping approach, we qualitatively accounted for 
recent experimental results for the damping of the quadrupole mode 
of a 2D Fermi gas. However, our results also showed that further work is needed to obtain a  quantitative understanding of the results. 

I am grateful to M.\ K\"ohl  for discussing the experimental results in Ref.~\cite{Vogt} with me.

\end{document}